    \documentstyle[epsf]{article} \catcode`\@=11
    \def\section{\@startsection{section}{1}{\z@}%
    {-3.5ex plus -1ex minus -.5ex}{1.5ex plus.3ex}{\bf }}
    \def\subsection{\@startsection{subsection}{1}{\z@}%
    {-3.5ex plus-1ex minus-.5ex}{1.5ex plus.3ex}{\bf }} 
    
\begin{document}
    \hfill\parbox{4.77cm}{\Large\centering Annalen\\der
    Physik\\[-.2\baselineskip] {\small \underline{\copyright\ Johann
    Ambrosius Barth 1998}}} \vspace{.75cm}\newline{\Large\bf
Electronic states in topologically disordered systems
    }\vspace{.4cm}\newline{\bf   
U. Grimm~$^{1}$, R.A. R\"omer~$^{1}$, and G. Schliecker~$^{2}$
    }\vspace{.4cm}\newline\small
$^1$~Institut f{\"u}r Physik, Technische Universit{\"a}t, D-09107
  Chemnitz, Germany\\ 
$^2$~Max-Planck-Institut f\"ur Physik komplexer Systeme,
N\"othnitzer Str. 38, D-01187 Dresden, Germany
    \vspace{.2cm}\newline 
    Received 6 October 1998, accepted 8 October 1998 by B.~Kramer
    \vspace{.4cm}\newline\begin{minipage}[h]{\textwidth}\baselineskip=10pt
    {\bf  Abstract.}
    Networks generated from Voronoi tessellations of space are
    prototypes for topologically disordered systems. In order to
    assess the effects of random coordination on the statistical
    properties of energy spectra, we analyze tight-binding models for
    planar topologically disordered systems.  To this end, the
    networks are generated by a simple topological model covering a
    wide range of naturally observed random structures.  We find that
    the energy-level-spacing distributions exhibit level repulsion,
    similar to the spacings in the energetically disordered Anderson
    model of localization in the metallic regime.
    \end{minipage}\vspace{.4cm} \newline {\bf  Keywords:}
Tessellations; Disorder; Random matrix theory; Spectral statistics
    \newline\vspace{.2cm} \normalsize

\section{Introduction}

In most natural structures disorder and random coordination are the
rule rather than an exception. Therefore the investigation of random
tessellations has become a subject of growing interest \cite{Review}.
In addition to the structures found in nature, the Voronoi
construction, a generalization of the Wigner-Seitz construction,
allows to generate tessellations from sets of arbitrarily distributed
points \cite{Okabe}. The random Voronoi tessellation has served as the
prototype of topologically disordered structures in physics
\cite{RandomVoronoi}.
The aim of this work is a first study of the influence of random
coordination on the properties of the electronic spectra in
topologically disordered structures. Similar structures are expected
in liquid metals or alloys. Here we analyze a tight-binding (TB) model
with neighbor hopping on certain two-dimensional (2D) random graphs.

\section{Tight-binding model on a random graph}

\begin{figure}
\centerline{\epsfxsize=0.7\textwidth\epsfbox{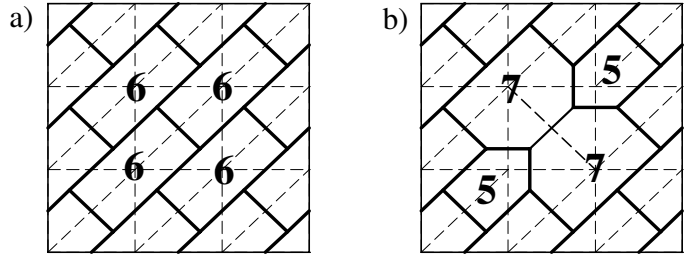}}
 \caption{\label{fig-structure}\small 
   a): Ordered $p=0$ triangulation (dashed thin lines) and mosaic
   (solid bold lines). The numbers indicate the number of edges of a
   closed loop (mosaic), or, equivalently, the number of neighbors of
   each vertex (triangulation). b): The same structure after a single
   flip.}
\end{figure}

A simple topological model introduced by Le Ca{\"e}r \cite{topmodI}
allows to generate efficiently random mosaics that are very similar to
Voronoi tessellations generated from disordered arrangements of
repulsive particles \cite{AirTable}.  Starting point of this model is
the triangular structure represented in Fig.~\ref{fig-structure}a
(dashed lines).  Disorder is achieved by random flips of the diagonal
bonds which are performed with probability $p$, yielding the most
disordered structures for $p=0.5$.  As an example, the central bond
has been flipped in Fig.~\ref{fig-structure}b. The dual graphs, the
mosaics, obtained by connection of neighboring triangles, are
represented as solid lines in Fig.~\ref{fig-structure}.

The TB Hamiltonian is given as
\begin{equation}
\label{eq-TightBindingHamiltonian}
H= \sum_{i,j=1}^{N} t_{i,j}\, c_i^\dagger c_j^{ } \ ,
\end{equation}
where $i,j$ label the sites of the 2D random graph. The total number
of sites is $N$, and $c_i^\dagger$ ($c_i$) denotes a fermionic
creation (annihilation) operator on site $i$ and we always use
periodic boundary conditions. The hopping matrix element between
neighboring sites $i$ and $j$ is $t_{i,j}=1$ and zero otherwise. The
eigenvalues and eigenstates of the TB Hamiltonian are calculated
employing standard diagonalization methods and Lanczos' algorithm.

\section{Density of States}

\begin{figure}
\centerline{\epsfxsize=\textwidth\epsfbox{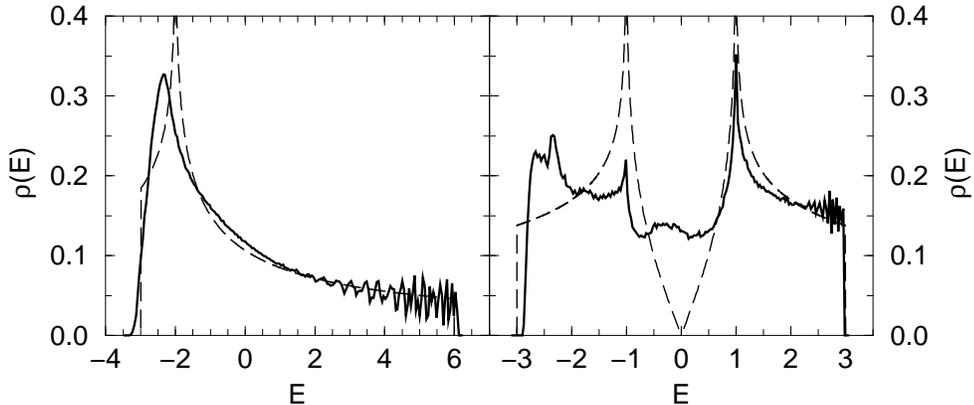}}
 \caption{\label{fig-dos}\small 
   Left: DOS for the ordered triangulation at $p=0$ (dashed line) and
   the disordered triangulation at $p=0.5$ (solid line). Right: Same
   as in the left panel but for the mosaic.}
\end{figure}

For the regular structures with $p=0$, the density of states (DOS) can
be obtained analytically. The dispersion relation of the triangular
structure reads
\begin{equation}
\epsilon_{\rm T}(\mbox{\boldmath $k$}) = 
2 \left[ \cos(k_x) + \cos(k_y) + \cos(k_x+k_y) \right],
\label{eq-epsilon-triangle}
\end{equation} 
where $\mbox{\boldmath $k$} = (k_x, k_y)$. In the infinite lattice,
the DOS is then given as
\begin{equation}
  \rho_{\rm T}(E) 
= 
\left( 4 \pi^2\right)^{-1} \int d\mbox{\boldmath $k$}\;
 \delta ( E-\epsilon_{\rm T}(\mbox{\boldmath $k$}) ) 
\label{eq-dos-triangle}
\end{equation}
where the integration is performed over the first Brillouin zone,
i.e., $k_x, k_y \in [-\pi, \pi)$. This integral can be expressed in
terms of complete elliptic integrals of the first kind.  The resulting
DOS is shown in Fig.\ \ref{fig-dos}. Note the logarithmic singularity
at $E=-2$.
For the mosaic structure at $p=0$, the dispersion relation is found to
be
\begin{equation}
  \label{eq-epsilon-hexagon}
  \epsilon_{\rm M}(\mbox{\boldmath $k$})= 
   \pm \sqrt{3 +\epsilon_{\rm T}(\mbox{\boldmath $k$}) }.
\end{equation}
The corresponding DOS
\begin{equation}
  \label{eq-dos-hexagon}
\rho_{\rm M}(E) 
=  |E|\, \rho_{\rm T} (E^2-3) 
\end{equation}
with logarithmic singularities at $E=\pm 1$ is also shown in Fig.\ 
\ref{fig-dos}. It is symmetric about $E=0$ because the ordered mosaic
is bipartite.

Fig.\ \ref{fig-dos} also shows the DOS of the most disordered
structures with $p=0.5$, obtained by averaging over $50$ realizations
on a system of $N=60\times 60$ sites. These DOS are rather smooth,
apart from oscillations at large energies which also show up in
finite-size data for the ordered structures.  The logarithmic
singularities are washed out by the disorder, and the negative-energy
peak for the mosaic structure almost disappears. However, the
positive-energy peak for the mosaic structure survives for $p=0.5$.

\section{Energy level distributions}

We now turn our attention to the statistical properties of the
eigenvalue spectrum of the Hamiltonian
(\ref{eq-TightBindingHamiltonian}).  Here, we focus on the
level-spacing distribution (LSD) $P(s)$ of the normalized energy
spacings $s$. In energetically disordered systems the metal-insulator
transition (MIT) in the 3D Anderson model of localization is
accompanied by a transition of the LSD. In the metallic regime, one
finds level repulsion with $P(s)= P_{\rm GOE}(s)$ of the Gaussian
orthogonal random matrix ensemble (GOE) \cite{mehta}, whereas in the
insulating regime, there is level clustering described by Poisson's
law $P_{\rm P}(s)=\exp(-s)$.  Level clustering is also expected for
ordered structures. For states at the MIT, an intermediate LSD has
been observed \cite{ZK}.

\begin{figure}
\centerline{\epsfxsize=\textwidth\epsfbox{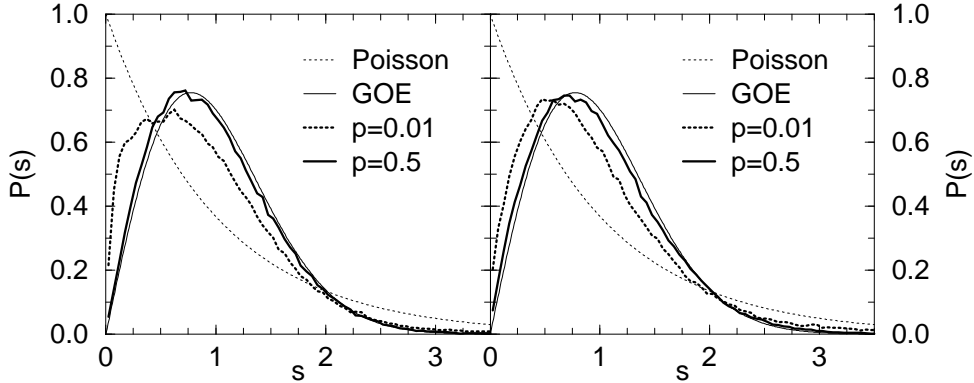}}
\caption{\label{fig-ps}\small 
  Left: $P(s)$ for the disordered triangulation at $p=0.01$ and
  $p=0.5$.  Thin lines indicate $P_{\rm GOE}(s)$ and $P_{\rm P}(s)$.
  Right: Same for the mosaic.}
\end{figure}

We have computed the LSD of the complete spectrum by properly
unfolding the spectrum \cite{HS} and averaging over $50$ realizations
on a system of $60\times 60$ sites.  In Fig.\ \ref{fig-ps}, we show
the results for random graphs with $p=0.01$ and $p=0.5$ in comparison
to $P_{\rm GOE}(s)$ and $P_{\rm P}(s)$. Both the mosaic and the
triangulation data show an enhanced level repulsion with increasing
$p$.  The topological structures for $p=0.01$ are still rather close to the
ordered system. Thus it is no surprise that the LSD in Fig.~3 still
shows reminiscences of level clustering. However, from an
investigation of the system size dependence we expect that even the
$p=0.01$ data tend towards the $P_{\rm GOE}(s)$ result for increasing
$N$.
We have also studied the behavior of other spectral correlations such
as the spectral rigidity $\Delta_3$ \cite{zhong}. Our results indicate
that there are still characteristic deviations from the GOE behavior
at the presently considered system sizes, although the system size
dependence again shows a tendency towards GOE. This is further
corroborated by an analysis of the statistical properties of the
eigenfunctions. Results will be published elsewhere.

\section{Looking at the eigenstates}

\begin{figure}
\centerline{
 {\epsfxsize=0.33\textwidth\epsfbox{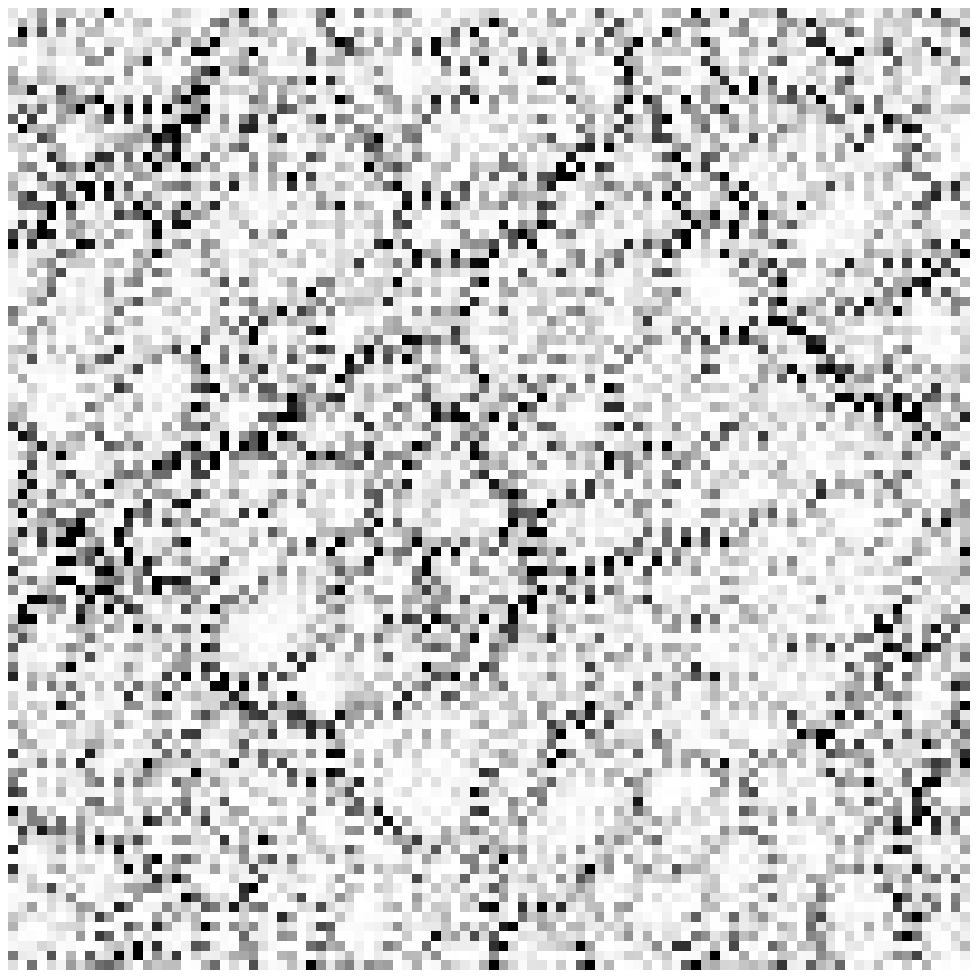}}
 {\epsfxsize=0.33\textwidth\epsfbox{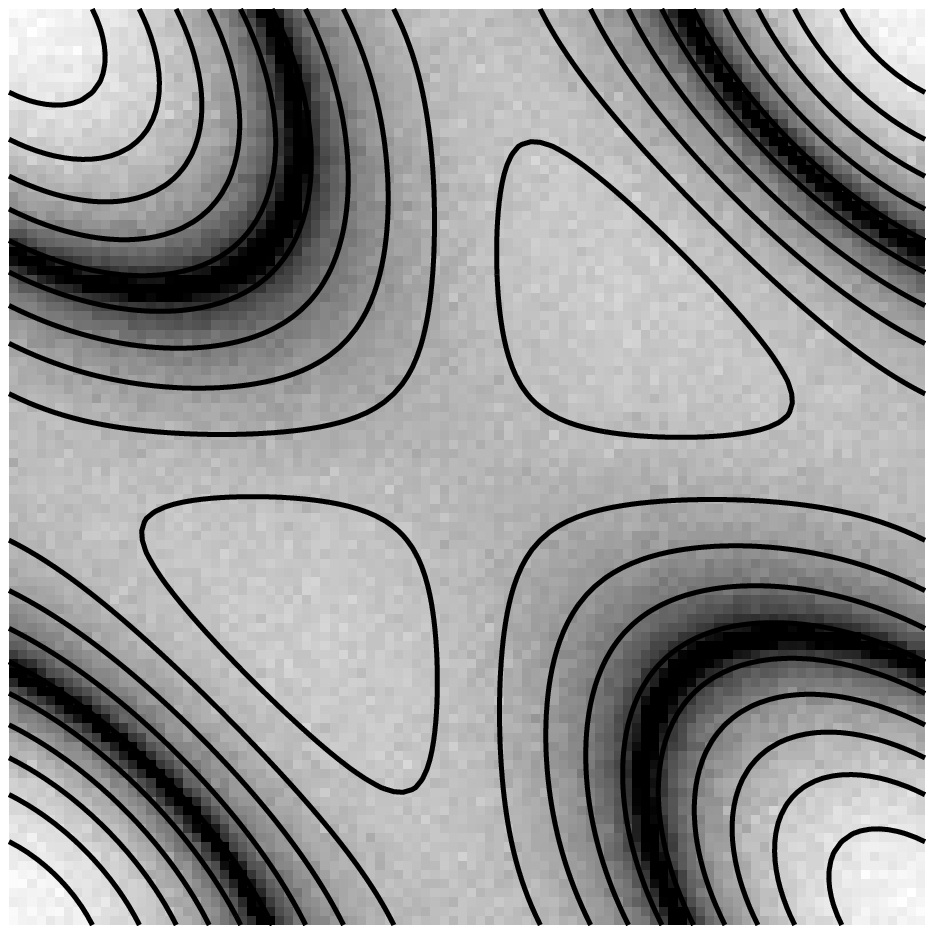}}
 {\epsfxsize=0.33\textwidth\epsfbox{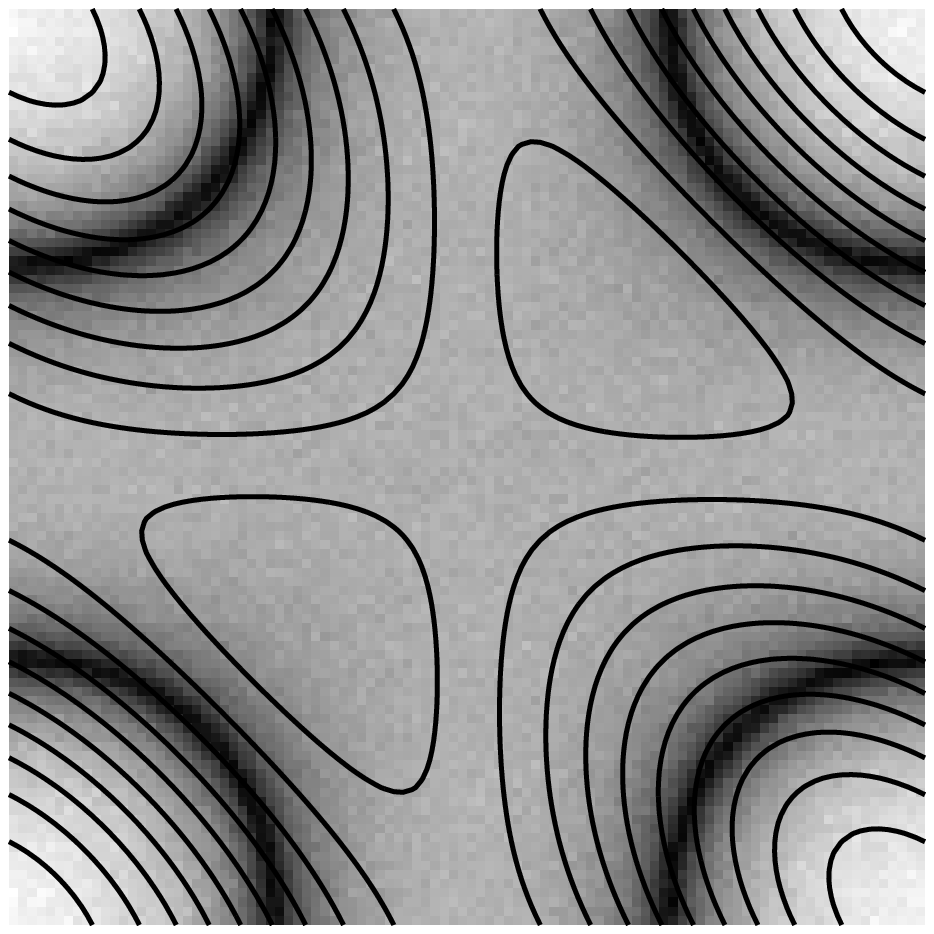}}
}
\caption{\label{fig-wf}\small 
  Left: Probability amplitude of a typical eigenstate at $E\approx
  1.5$ in the random triangulation with $p=0.5$ and $N= 100\times
  100$. Dark regions correspond to high probability amplitudes.
  Center: Averaged probability amplitude of Fourier-transformed
  eigenstates at $p=0.01$ with $E$ and $N$ as in the left panel. The
  lines indicate the Fermi surfaces of the ordered structure. Right:
  Same as in the center but with $p=0.5$. }
\end{figure}

In Fig.\ \ref{fig-wf} we show the probability density of a typical
eigenfunction $\psi (\mbox{\boldmath $r$})$ for the triangular
structure at energy $E\approx 1.5$ close to the center of the band. In
agreement with the results presented above, we find that the
eigenfunction extends over the whole system. This state is of course
not a Bloch state, but nevertheless may be viewed as a superposition
of eigenstates of the ordered system at energies $E' = E \pm \Delta
E$, where $\Delta E$ represents an energy level broadening due to the
disorder. An eigenstate of the weakly disordered system in Fourier
space should therefore exhibit a broadened Fermi surface when
interpreting $E$ as the Fermi energy \cite{eilmes}.  We also plot the
probability density of the Fourier-transformed eigenstate
$\tilde{\psi}(\mbox{\boldmath $k$})$ in Fig.\ \ref{fig-wf} and compare
it to the sharp Fermi surface of the ordered structure.  The latter
can easily be computed from the dispersion relation
(\ref{eq-epsilon-triangle}).  For the strongly disordered case at
$p=0.5$ we find an almost four-fold symmetry, whereas the ordered
triangulated structure only has a reflection symmetry with respect to
the line $k_x=-k_y$. This is to be expected because for $p=0.5$ both
orientations for the diagonals (cp.\ Fig.\ \ref{fig-structure}) are
equally probable. Thus the four-fold symmetry of the square lattice is
restored on average.
Similar results hold for the mosaic structures.

\section{Conclusions}

We have considered the spectral properties of a TB Hamiltonian defined
on certain 2D random graphs. Although the systems are not
energetically disordered, the LSD exhibits level repulsion much as for
the energetically disordered Anderson model of localization in the
metallic regime \cite{HS} and also for TB models defined on
quasiperiodic tilings \cite{zhong}. However, more evidence, such as
the spectral rigidity mentioned above, is needed before we can
conclusively decide whether the level repulsion is indeed identical to
the behavior of GOE. We remark that preliminary results for
eigenfunction statistics may indicate systematic deviations from the
GOE results for the most disordered mosaic structures.
    \vspace{0.6cm}\newline{\small 
      We thank F.\ Milde for programming help in the initial stages
      of this project.
    }

\begin{thebibliography}{99} \itemsep-2pt \small\frenchspacing
%
\bibitem{Review}
D. Weaire and N. Rivier, Contemp. Phys. {\bf 25} (1984) 59;
J. A. Glazier and D. Weaire, J. Phys. C {\bf 4} (1992) 1867;
N. Rivier, in {\it Disorder and Granular Media}, D. Bideau and A. Hansen
 (eds.), Elsevier Science Publishers B.~V.~, (1993) 55;
J. Stavans, Rep. Prog. Phys. {\bf 56} (1993) 733     
\bibitem{Okabe} A. Okabe, B. Boots, and K. Sughihara, {\it Spatial
    Tessellations}, Wiley, Chichester (1995)
\bibitem{RandomVoronoi}
N. H. Christ, R. Friedberg and T. D. Lee, Nucl. Phys. B 
{\bf 202} (1982) 89;
J. M. Drouffe and C. Itzykson, Nucl. Phys. B {\bf 235} (1984) 45;
P. H. Winterfield, L. E. Scriven, and M. T. Davis, 
J.~Phys.~C {\bf 14} (1981) 2361
\bibitem{topmodI}
G. Le Ca{\"e}r, J. Phys. A {\bf 24} (1991) 4655 
\bibitem{AirTable} 
J. Lema{\^\i}tre, A. Gervois, J. P. Troadec, N.
  Rivier, M. Ammi, L. Oger, and D. Bideau, Phil. Mag. B {\bf 67}
  (1993) 347

\bibitem{mehta} M.~L.~Mehta, {\it Random Matrices}, 2nd ed.\ Academic
  Press, Boston (1990)

\bibitem{ZK} I.~K.~Zharekeshev and B.~Kramer, Phys.\ Rev.\ Lett.\ {\bf
    79} (1997) 717

\bibitem{HS} E.~Hofstetter and M.~Schreiber, Phys.\ Rev.\ B {\bf 48}
  (1993) 16979; {\bf 49} (1994) 14726; Phys.\ Rev.\ Lett.\ {\bf 73}
  (1994) 3137

\bibitem{zhong} J. X.  Zhong, U. Grimm, R. A. R\"{o}mer, and M.
  Schreiber, Phys.  Rev. Lett. {\bf 80} (1998) 3996 

\bibitem{eilmes} A. Eilmes, R. A. R\"{o}mer, and M. Schreiber, Eur.
  Phys. J. B {\bf 1} (1998) 29


    \end{thebibliography}
    \end{document}